\documentclass[preprint,showpacs,preprintnumbers,amsmath,amssymb]{revtex4}
\usepackage{epsfig,amsmath,amssymb,graphics,color,calc,wasysym}

\newcommand{\be}{\begin{equation}}
\newcommand{\ee}{\end{equation}}
\newcommand{\ba}{\begin{eqnarray}}
\newcommand{\ea}{\end{eqnarray}}

\renewcommand{\phi}{\varphi}

\begin{document}

\title{Non-Gaussian Dynamics in Smectic Liquid Crystals of Parallel Hard Rods}

\author{Rik Matena}
\author{Marjolein Dijkstra}
\author{Alessandro Patti\footnote{Electronic mail: a.patti@uu.nl and m.dijkstra1@uu.nl}}
\affiliation{Soft Condensed Matter, Debye Institute for NanoMaterials Science, Utrecht University, Princetonplein 5, 3584 CC, Utrecht, The Netherlands}

\begin{abstract}
Using computer simulations, we studied the diffusion and
structural  relaxation in \textit{equilibrium} smectic liquid
crystal bulk phases of parallel hard spherocylinders. These
systems exhibit a non-Gaussian  layer-to-layer diffusion due to
the presence of periodic barriers and transient cages, and show
remarkable similarities with the behavior of
\textit{out-of-equilibrium} supercooled liquids. We detect a very
slow inter-layer relaxation dynamics over the whole density range
of the stable  smectic phase which spans a time interval of four
time decades. The intrinsic nature of the layered structure yields
a hopping-type diffusion which becomes more heterogeneous for
higher packing fractions. In contrast, the in-layer dynamics is
typical of a dense fluid with a relatively fast decay. Our results
on the dynamic behavior agree well  with that observed in systems
of freely rotating hard rods, but differ quantitavely, as the height of the periodic barriers reduces to zero at the nematic-smectic transition for aligned rods, while it remains finite for freely rotating rods.
\end{abstract}


\maketitle

\section{I. Introduction}
\label{sec:Introduction}

Non-Gaussian diffusion and dynamical heterogeneities have been
shown to slow down  or completely arrest the structural relaxation
of  systems, such as supercooled liquids \cite{berthier, kob,
brambilla} and gels \cite{abete, puertas}. The heterogeneous
dynamics of these systems, in which individual particles are
trapped in transient cages by neighboring particles, explains the
non-exponential relaxation and non-Gaussian diffusive behavior.
Two scenarios are usually proposed to explain the non-exponential
relaxation behavior: a \textit{heterogeneous scenario} in which
the particles relax exponentially at different relaxation rates,
and a \textit{homogeneous scenario}  where the particles relax
non-exponentially at nearly identical rates \cite{richert}.
Recently, the analysis of dynamical heterogeneities has been
extended to other complex systems, such as granular media
\cite{granular} and liquid crystals (LCs) in confined nanopores
\cite{lefort, ji}. Interestingly, the dynamics of LCs has been
shown to share many features with supercooled liquids, especially
in the so-called LC isotropic phase \cite{cang, gottke} and in the
smectic LC phase \cite {patti}. The LC isotropic phase is a
macroscopically homogeneous liquid phase, which exhibits nematic
ordered domains near the isotropic-nematic (IN) transition. The
size of these nematic domains, which are driven by a precursor of
the nematic phase, increases upon approaching the IN-transition
\cite{degennes}. The caging behavior, i.e., the temporary
localization of individual particles, as observed in supercooled
liquids, is  caused by the transient structural inhomogeneities of
the nematic domains in the isotropic LC phase. Hence, the
heterogeneous dynamics of such an isotropic  phase with
pre-nematic-order resembles that of supercooled liquids
\cite{cang}. The non-Gaussian behavior becomes even stronger  by
confining the LC in a nanoporous material \cite{lefort, ji}. The
reason is that confinement modifies the dynamics of the LC
particles, which yields a non-uniform relaxation  depending on
e.g., the distance to the pore surface and the pore size, which is
indeed  observed experimentally by dielectric spectroscopy
\cite{frunza} and quasi-elastic neutron scattering \cite{lefort,
guegan}.

For the bulk smectic LCs,  recent developments of  experimental
techniques (e.g. NMR coupled to strong magnetic field gradients
\cite{furo}, or fluorescent labeling of rods \cite{lettinga}),
allowed for direct observations of non-Gaussian quasi-quantized
layer-to-layer diffusion. In the light of these advances, Bier
\textit{et al.} proposed a dynamic density functional approach to
study the self-diffusion in colloidal dispersions of infinitely
elongated particles \cite{bier}. In particular, they investigated
the effect of the      local fluid structure, which temporarily
cages  individual particles and competes with the  one-dimensional
"permanent" barriers due to the smectic layered structure. This
theoretical work shows qualitative agreement with recent
experiments on the self-diffusion of filamentous bacteriophage
\textit{fd} viruses  through smectic layers \cite{lettinga}. In
both studies, the self-part of the van Hove correlation function
showed clear evidence of an inter-layer diffusion (or
\textit{permeation}) occurring by discontinuous jumps of nearly
one rod length. Simulations on freely rotating hard rods confirmed
the important role of temporary cages and permanent barriers on
the non-Gaussian permeation through smectic layers, and revealed
new insights on the relaxation behavior and cooperative motion of
stringlike clusters  \cite{patti}.

In this work, we investigate the  diffusion in bulk smectic liquid
crystals of perfectly aligned hard spherocylinders. The phase
diagram of parallel hard spherocylinders exhibit
nematic-to-smectic (N-Sm) and smectic-to-crystal (Sm-K) phase
transitions in a broad range of length-to-diameter ratios of the
rods \cite{stroobants, veerman}. Using computer simulations, we
are able to study the long-time relaxation decay which
approximately spans up to four time decades in the whole density
range from the N-Sm to the Sm-K transition. We investigate the
substantial differences between the fluid-like in-layer and the
hopping-type inter-layer dynamics, which are  caused by the
temporary cages and permanent barriers. In addition,  we study the
heterogeneous character of the layer-to-layer dynamics, which is
not exclusively a feature of confined LCs, and we determine the
crossover between the cage regime, with the particles rattling
around their center of mass, and the long-time diffusive regime as
a function of density.

\section{II. Model and Simulations}
\label{sec:Model and Simulations}

Our system is composed of $N=2100$ parallel hard spherocylinders
with aspect ratio $L^{*}\equiv L/D=5$, where $D$ is the diameter
of the hemispherical caps joined together by a cylindrical part of
length $L$. Hence, the overall length of the rods is $L+D$. The
phase diagram of this system displays stable nematic, smectic, and
crystal phases \cite{veerman}. For $L^{*}=5$, the smectic phase
melts into a nematic phase for $P^{*} \lesssim 2.1$, and
crystallizes for $P^{*}\gtrsim 6.3$, where $P^{*}\equiv
Pv_{0}/k_{B}T$ is the reduced pressure with $k_{B}$ Boltzmann's
constant and $v_{0}=\pi (D^3/6 + L D^2/4)$ the molecular volume.
We studied the dynamics of this system at  pressures $P^{*}=2.0$,
2.5, 3.0, 4.0, and 5.0 corresponding to packing fractions
$\eta=Nv_{0}/V=0.394$, 0.441, 0.475, 0.525, and 0.563,
respectively. We performed Monte Carlo (MC) simulation in a
rectangular box of volume  $V$  with 7 smectic layers of aligned
rods and we employ periodic boundary conditions.

We first performed equilibration runs using MC simulations in the
isobaric-isothermal (NPT) ensemble. The initial configurations for
the smectic phase were obtained  by expanding a solid phase at
pressures $P^{*}=2.0$, 2.5, 3.0, and 4.0. The smectic phase at
$P^{*}=5.0$ was obtained by a compression run from $P^{*}=4.0$. We
only performed translational moves, which were accepted if no
overlap was detected, whereas rotational moves were not allowed.
Volume changes have been attempted every \textit{N} MC cycles by
randomly changing the three box lengths independently. The
systems were considered to be equilibrated when the packing
fraction had reached a constant value. In the production runs, we
carried out MC simulations in the canonical (NVT) ensemble, i.e.,
we kept the volume constant, as the  collective moves associated
with the volume changes would not correspond with Brownian
dynamics.   The maximum displacement of the MC moves was chosen in
 such a way that we achieve (\textit{i}) a reasonable time of
simulation, (\textit{ii}) a satisfactory acceptance rate, and
(\textit{iii}) a suitable description of the Brownian motion of
the particles in a colloidal suspension. To this end, we
monitored the mean square displacement in the \textit{z}- and
\textit{xy}-directions for several values of the maximum step size
$\delta_{max}$, with $\delta_{max, z}=2\delta_{max, xy}$ due to
the anisotropy of the short-time self-diffusion coefficient of the
rods \cite{Doi}. $\delta_{max, xy}=D/10$ and $\delta_{max, z}=D/5$
were found to be the optimal values satisfying the above
requirements. We have neglected the hydrodynamic effects, as it
was shown recently by computer simulations of highly concentrated
rod suspensions, that the dynamics is dominated primarily by the
excluded volume and steric effects \cite{pryamitsyn}.

As unit of time, we have chosen  $\tau\equiv D^{2}/D_{tr}$, where
$D_{tr}$ is the translational short-time diffusion coefficient,
which is the isotropic average of the diffusion coefficients in the
three space dimensions: $D_{tr} \equiv
(D_{\parallel}+2D_{\perp})/3$.

\section{III. Results}
\label{sec:Results}

The intrinsic nature of the smectic phases  can be appreciated by
computing the in-layer $g_{\perp}(x, y)$, and inter-layer
$g_{\parallel}(z)$ pair correlation functions as a function of $r_{\perp} = \sqrt{x^2 + y^2}$ and $z$, respectively. In Fig. 1, we
display $g_{\perp}(x, y)$ and $g_{\parallel}(z)$  for varying
pressures. We find features of fluid-like behavior for
$g_{\perp}(x, y)$: the first peak is located at a distance
approximately equal to one diameter length, and an exponential
decay of the  oscillations to one at long distances.  By contrast,
$g_{\parallel}(z)$ shows pronounced periodic correlations which
reveal the layered structure along the nematic director
\textit{\^{n}}. The location of the peaks corresponds to the
center of the smectic layers, where the particle density is
maximal. In Fig. 2, we show the top and side view of two typical
configurations of the smectic LC phase at $P^{*}=2.5$ and 5.0,
where one can clearly see  the periodic structure of the smectic
layers and the two-dimensional fluid-like structure within each
layer.
\begin{figure}[!h]
\center
\includegraphics[width=0.48\textwidth]{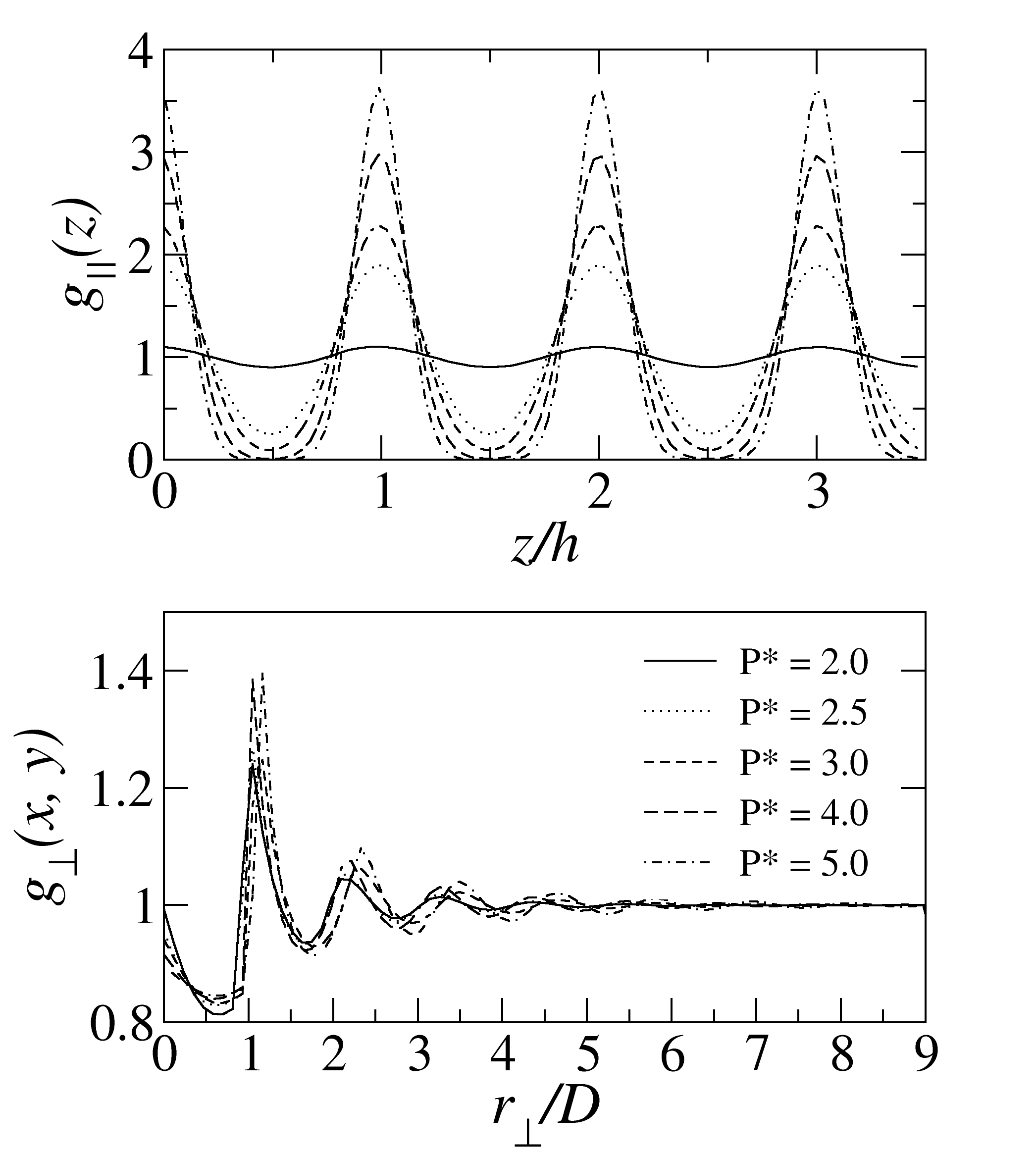}
\caption{(a) Inter-layer $g_{\parallel}(z)$, and (b) in-layer
$g_{\perp}(x, y)$ pair correlation functions as a function of $z$ and $r_{\perp} = \sqrt{x^2 + y^2}$, respectively, for varying pressures
as labeled.}
\end{figure}

Upon increasing the pressure, the peaks become more pronounced
indicating that the smectic layers are more defined and the
layer spacing reduces significantly from $h=7.03D$ at
$P^{*}=2.0$ to $h=6.48D$ at $P^{*}=5.0$. At $P^{*}=2.0$, very
close to the N-Sm transition, it is difficult to distinguish the
individual layers as the amplitude of the periodic structure
reduces continuously, and $g_{\parallel}(z)$ approaches a nearly
flat profile in the proximity of the continuous N-Sm transition. 
At pressures $P^{*}\geq 4.0$, the peaks become more
pronounced, while the minima, corresponding to the interlayer
spacings tends to zero. Hence, it becomes more difficult for the
particles to diffuse from layer to layer at these high pressures.

\begin{figure}[!h]
\center
\includegraphics[width=0.48\textwidth]{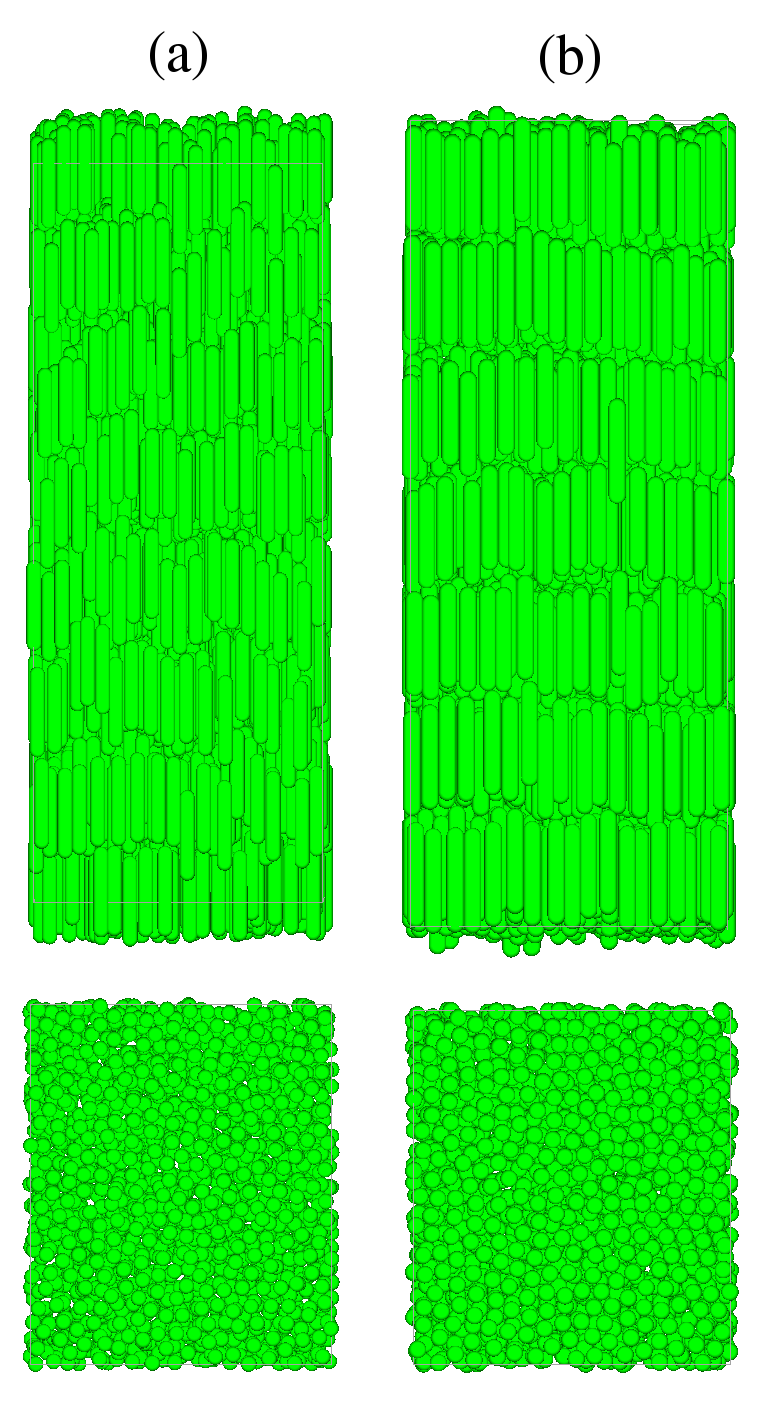}
\caption{Top and side views of two bulk smectic liquid crystal
phases observed at (a) $P^{*}=2.5$  and (b) $P^{*}=5.0$.}
\end{figure}

In addition,  we measure the (relative) probability $\pi(z)$ of
finding a particle at position $z$ with the $z$-axis chosen parallel to
the nematic director $\hat{n}$.  We estimated the potential energy
barrier  from  the Boltzmann factor $\pi(z) \propto
\exp(-U(z)/k_{B}T$) as in Ref. \cite{lettinga}. $U(z)$
denotes the effective potential and quantifies the potential energy
barrier for the layer-to-layer diffusion. In Fig. 3, we give
$U(z)$ at different pressures with the fitting function
$U(z)=\sum_{i=1}^{n} U_{i}[\sin(\pi z/h)]^{2i}$, where
$U_{0}=\sum_{i=1}^{n}U_{i}$ is the potential barrier height and
$h$ the inter-layer spacing. We find $U_{0}=0.2k_{B}T$  at
$P^{*}=2.0$ and $U_{0}=8.3k_{B}T$ at $P^{*}=5.0$. The potential
barrier for a rod to diffuse from layer to layer increases with
increasing packing fraction, and becomes less steep when the N-Sm
phase transition is approached at lower densities.

\begin{figure}[!h]
\center
\includegraphics[width=0.48\textwidth]{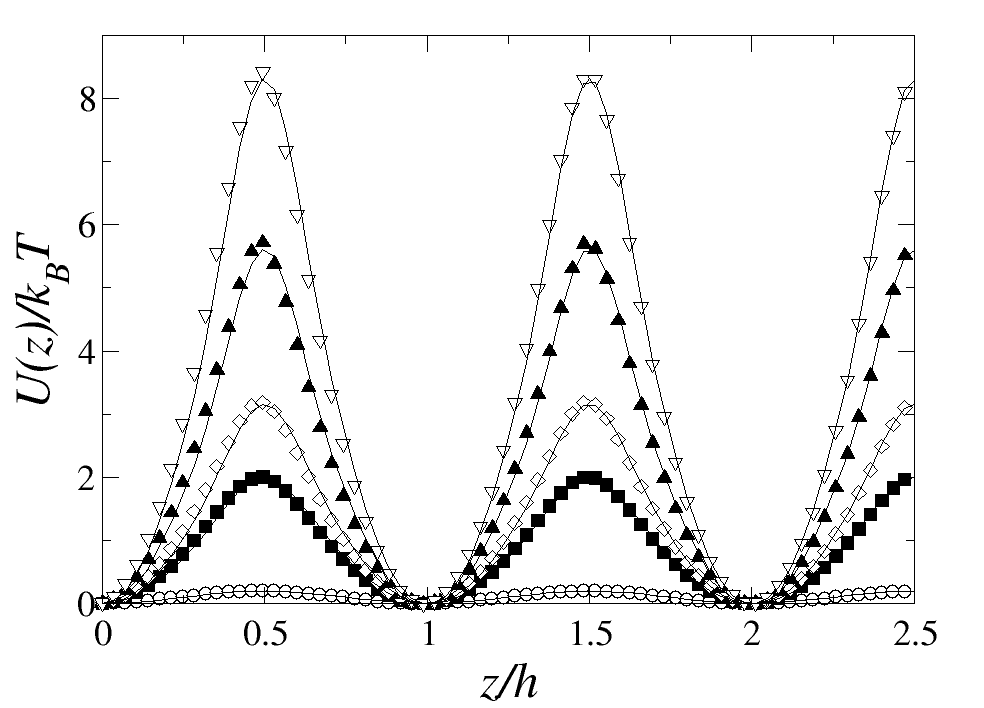}
\caption{Effective potential $U(z)$ in the bulk smectic phase at $P^{*}=2.0$ ($\Circle$), 2.5 ($\blacksquare$), 3.0 ($\Diamond$), 4.0 ($\blacktriangle$), and 5.0 ($\triangledown$). The solid lines are fits.}
\end{figure}

These results should be compared with those of Ref.
\cite{vanduijneveldt} and the barrier heights estimated recently
for  smectic phases of freely-rotating hard rods with $L^{*}=5$ 
\cite{patti}. Interestingly, we find that the barrier height  in
smectic LC phases is higher for aligned hard rods than for freely
rotating rods at the same packing fraction, i.e., we find 
$U_{0}=5.1k_{B}T$ at $\eta=0.508$ ($P^{*}=3.7$) for aligned rods, which should be
compared with $U_{0}=3.5 k_B T$ for freely rotating rods, and $U_{0}=7.8k_{B}T$ at
$\eta=0.557$ ($P^{*}=4.8$) for parallel rods to be compared with $U_{0}=7.5k_{B}T$.
The difference in barrier height decreases with  increasing
pressure, and we expect that it tends to zero upon approaching the
Sm-K transition, where the freely rotating rods become more and
more aligned. As the potential barriers are significantly lower
for freely rotating rods at low packing fractions, we conclude
that  rotational degrees of freedom  facilitate significantly the
layer-to-layer diffusion.

Barrier-free diffusion pathways can be observed  even at high
densities as a result of screw dislocations. Such structural
defects create helical connections between neighboring smectic
layers where the rods diffuse as  in the nematic phase
\cite{selinger}. It is not an easy task to model screw
dislocations in computer simulations, as these are not  compatible
with  periodic boundary conditions. Slip boundary conditions may 
overcome this complication \cite{lill}.

\begin{figure}[!h]
\center
\includegraphics[width=0.48\textwidth]{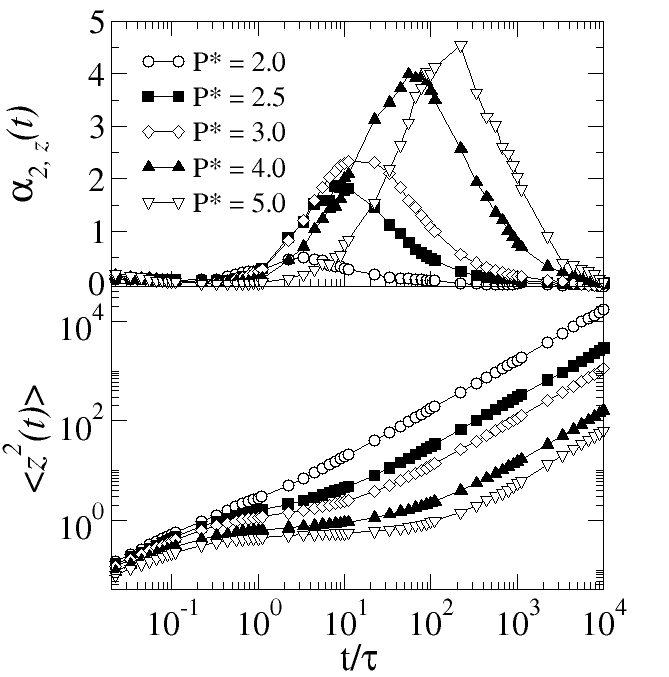}
\caption{Non-Gaussian parameter $\alpha_{2,
\textit{z}}(\textit{t})$ and mean square displacement in the
direction parallel to the nematic director, calculated at the
pressures indicated in the top frame.}
\end{figure}

To study the non-Gaussian behavior  of the layer-to-layer
diffusion, we computed the following non-Gaussian parameter \cite{rahman}:
\begin{equation}
\alpha_{2,z}(t)=\dfrac{ \langle\Delta
z(t)^{4}\rangle}{3\langle\Delta z(t)^{2}\rangle^{2}}-1
\end{equation}
where $\Delta z(t)=z(t_{0}+t)-z(t_{0})$ is the displacement of the
rods in the \textit{z}-direction in the time interval starting at
$t_{0}$ and ending at $t_{0}+t$, and $\langle...\rangle$ denotes
an ensemble average over all particles and initial time $t_{0}$.
For the in-layer diffusion, a similar non-Gaussian parameter,
$\alpha_{2,xy}(t)$, can be computed. In Fig. 4, we show
$\alpha_{2,z}(t)$ at different pressures. This parameter
quantifies the deviation from the Gaussian behavior of the
probability density function for  single-particle diffusion. At
short times, the particles are freely diffusing as the trapping
cage formed by  the surrounding neighbors, is not reached yet, and
hence, $\alpha_{2, z}$ is basically zero. At long times, the
system enters the diffusive regime and the non-Gaussian parameter
tends to zero. At intermediate times, the motion of each particle
is hampered by its neighbors and becomes sub-diffusive. In this
time interval, $\alpha_{2, z}$ is non-vanishing, indicating the
development of dynamical heterogeneities. Additionally, we find
that the peak height of $\alpha_{2, z}(t^*)$ at $t^*$ increases
and moves to larger values of $t^*$ upon increasing the packing
fraction, as the time for the particles to escape out of their
cage increases. We also plot the mean square displacement (MSD)
$\langle z^2(t) \rangle$ in Fig. 4 for the same state points. The
MSD show a clear cage-trapping plateau, which becomes more
pronounced upon increasing pressure.  The time $t^*$ at which
$\alpha_{2, z}$ displays a maximum corresponds to the  end of the
plateau observed in  the MSD as can be observed clearly in Fig. 4.

Interestingly, non-Gaussian dynamics due to cage-escape processes
have also been observed in single-particle diffusion in periodic
external potentials \cite{vorselaars}, 2D liquids
\cite{hurley}, cluster crystals \cite{likos}, but also in
glasses  \cite{donati, weeks, weeks2, kegel}. In particular, it
was shown that the time to escape out of a  cage increases because
cage rearrangement (or recaging) involves a larger number of
particles \cite{donati, weeks, weeks2, kegel} upon approaching the
glass transition. $\alpha_{2, xy}(t)$ (not shown here) does not
deviate significantly from zero. At the highest pressure studied,
the peak is lower than 0.2, confirming that the in-layer
relaxation dynamics is (nearly) diffusive. The behavior of the non-Gaussian
parameter $\alpha_{2, z}$ is consistent with the theoretical
predictions in systems of infinitely long parallel hard rods
\cite{bier}, and with the simulation results of freely rotating
hard rods \cite{patti}, although the latter exhibit higher peaks
especially close to the smectic-to-crystal transition. Stronger 
deviations have been observed in colloidal systems
with short-range attractions when approaching the gel transition
\cite{puertas2}, or in permanent gels where \textit{static}
heterogeneities give rise to a plateau at long times (i.e., the
non-Gaussian parameter does not vanish) \cite{abete, abete2}.
Glass transitions, by contrast, are usually characterized by
weaker deviations \cite{kegel, weeks}.

The periodic shape of the effective potential  implies a
\textit{hopping-type} diffusion in the direction of $\hat{n}$,
with the rods rattling around in a given layer until they overcome
the free-energy barrier shown in  Fig. 3 and jump  to a neighboring layer. We quantify this layer-to-layer
diffusion by calculating the self part of the Van Hove correlation
function (VHF) \cite{hansen}, defined as:
\begin{equation}
G_{s}(\textit{z},\textit{t})=\frac{1}{N}\left\langle \sum_{i=1}^N \delta\left[\textit{z}-(\textit{z}_{i}(t_{0}+t)-\textit{z}_{i}(t_{0}))\right]\right\rangle
\end{equation}
where $\delta$ is the Dirac delta function.
$G_{s}(\textit{z},\textit{t})$ measures the probability
distribution  for the $z$-displacements of the rods at time
$t_{0}+t$, given their $z$-positions at $t_{0}$. In Fig. 5, the
self VHF is presented as a function of $z$ at $P^{*}=2.0$ to 5.0.
At $P^{*}=3.0$ to 5.0, we observe the appearance of peaks at
distances that correspond to the center of the smectic layers in
the \textit{z} direction. No peaks are observed at $P^{*}=2.0$,
where the  barrier height ($0.2k_{B}T$) of the effective potential
for the layer-to-layer diffusion is sufficiently small, that there
is no hopping-type diffusion  between  neighboring layers. As a
result, the discontinuous diffusion observed in strong smectic
phases, with the particles occupying quasi-discretized positions,
is substituted by a quasi-continuous Gaussian diffusion for weak
smectic phases. At long times, less and less particles are still
at their original positions, and consequently, the profiles of the
VHF become almost constant in a nematic phase or periodically
peaked in a smectic phase. The presence of peaks at  distances of
neighboring layers for small to intermediate pressures show that a
significant number of particles have displaced a long distance
even at short times. These \textit{fast}-moving particles
contribute to the heterogeneous dynamics of the system and affect
its structural relaxation. In recent simulations of smectic phases
of freely rotating hard rods, it was shown that the fast-moving
particles form stringlike clusters which exhibit cooperative
motion \cite{patti}. We believe that this result is still valid if
the orientational degrees of freedom are frozen out, but further
investigation is needed to address this point in more detail.

\begin{figure}[!h]
\center
\includegraphics[width=0.48\textwidth]{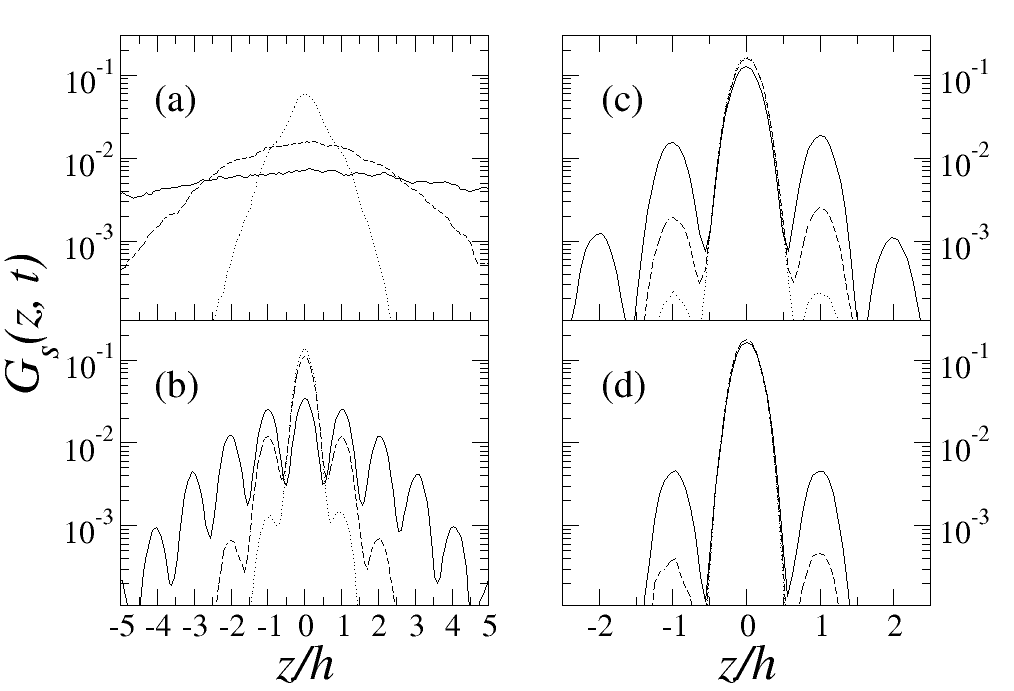}
\caption{Self part of the Van Hove function
$G_{s}(\textit{z},\textit{t})$ at $t/\tau=10$ (dotted lines),
$t/\tau=100$ (dashed lines), and $t/\tau=1000$ (solid lines) for
pressure $P^*=2.0$ (a), 3.0 (b), 4.0 (c), and 5.0 (d).}
\end{figure}
Finally, we also study the structural relaxation by calculating
the self-part of the intermediate scattering function:
\begin{equation}
F_{s}(t)=<\exp[i\textbf{q}\Delta \textbf{r}(t)]>
\end{equation}
at wave vectors $\textbf{q}D=(0,0,q_{z})$ and $(q_{x},q_{y},0)$,
with $q_{z}\simeq1$ and $(q^{2}_{x}+q^{2}_{y})^{1/2}\simeq6$,
which correspond to the main peaks of the static structure factor.
$\Delta \textbf{r}(t)$ is the displacement of a particle in the
time interval $t$. Results at pressures $P^{*}=$ 2.0, 3.0, and 4.0
are shown in Fig. 6.
For the pressure range $P^{*}=2.5-5.0$, the inter-layer dynamics  shows a
significantly slow relaxation which is characterized by a two-step
decay. In the first step, which is relatively fast, each rod
rattles around its original position without feeling the presence
of the surrounding neighbors. We detect an exponential decay of
$F_{s, z}(t)$ towards a plateau whose height and time extension
increase with pressure, as also observed for colloidal glasses
\cite{brambilla}. The plateau establishes the beginning of the
cage regime, where the particles start to interact with their
nearest neighbors which form a temporary cage for the particle.
The following step takes place at much longer times (note the
logarithmic scale of Fig. 6) and marks the escape from the cage
regime. The second decay of $F_{s, z}$ is well fitted by a
stretched exponential function of the form $\exp[−(t/tr
)^{\beta}]$, where $\beta \simeq 0.6$ and $t_{r}$ is the time at
which the intermediate scattering function decays to $e^{-1}$. The
stretched exponential form of the relaxation decay at long times
confirms the heterogeneous nature of the inter layer dynamics. In
Fig. 7, the relaxation time is given as a function of the packing
fraction and fitted with a power law covering 4 time decades in a
density interval spanning from the N-Sm to the Sm-K transition. At
$P^{*}=2.0$, we still observe an initial exponential decay at
short times and a stretched exponential decay at long times, but
it is very hard to detect a plateau (if any) due to the weak
smectic character of the bulk phase and, hence, to the weak
permanent background barriers (see Fig. 3). Nevertheless, there
might be a discontinuity between two separated relaxation regimes,
since $F_{s, z}$ cannot be fitted by a single scaling law.

\begin{figure}[!h]
\center
\includegraphics[width=0.48\textwidth]{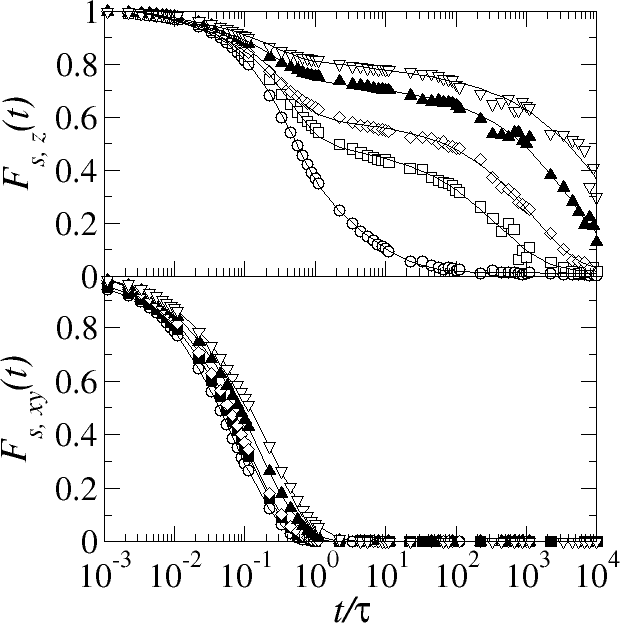}
\caption{Self-intermediate scattering function $F_s(t)$ for the
structural relaxation in the \textit{z} (top) and \textit{xy}
(bottom) direction at $P^{*}=2.0$ ($\Circle$), 2.5
($\blacksquare$), 3.0 ($\Diamond$), 4.0 ($\blacktriangle$), and
5.0 ($\triangledown$). The solid lines are fits.}
\end{figure}

By contrast, the in-layer relaxation is very fast and occurs  in a
single step. As can be observed in Fig. 7, the relaxation time, which covers basically  half a time decade, 
does not change significantly as a function of density. Interestingly, the long-time decay
of $F_{s, xy}$, which is exponential at short times, is also
fitted by a stretched exponential function with $\beta \simeq
0.7$ at long times. This behavior is unusual for simple fluids, where a single
exponential decay is  observed, and can also not be associated with that of a supercooled liquid as the
characteristic cage-trapping plateau is absent. However, the in-layer dynamics in smectic liquid
crystals is similar to that of low-density supercooled liquids
which do not display a plateau, but only a stretched
exponential relaxation at long times \cite{elmasri}.

\begin{figure}[!h]
\center
\includegraphics[width=0.48\textwidth]{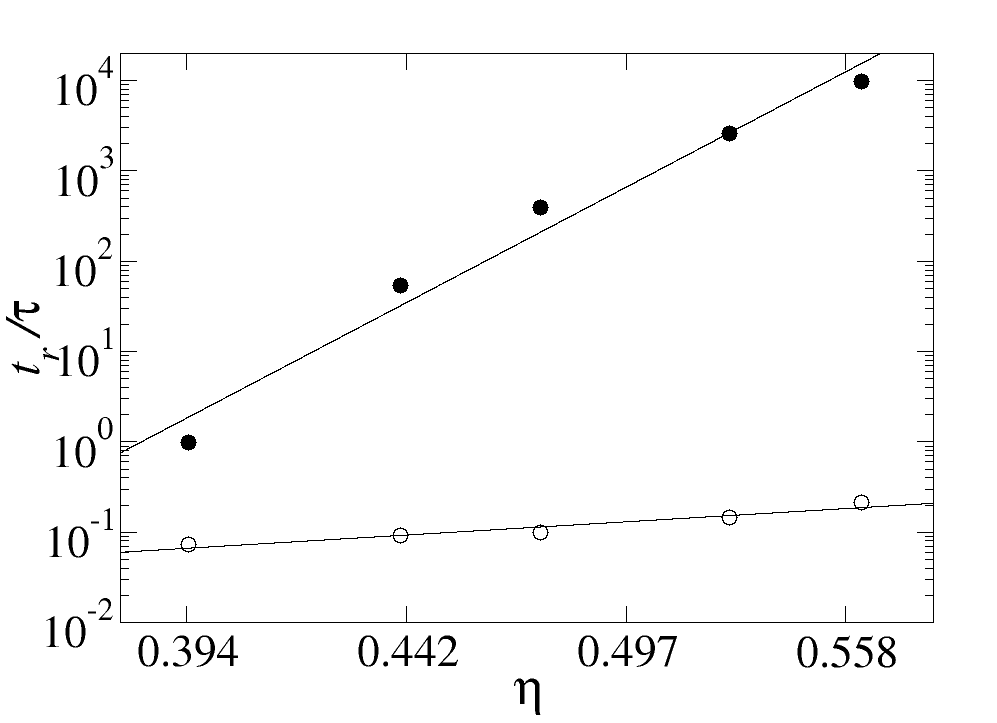}
\caption{Relaxation time $t_r/\tau$ as a function of  packing fraction $\eta$. The solid and empty circles refer to the inter-layer and in-layer relaxation, respectively. The solid lines are power-law fits.}
\end{figure}

\section{IV. Conclusions}
\label{sec:Conclusions}

In conclusion, we have studied using computer simulations the diffusion  of perfectly aligned rodlike particles in smectic liquid crystal phases, where the inter-layer dynamics exhibits a non-Gaussian rattling-and-jumping-type diffusion due to the  simultaneous presence of temporary cages and permanent barriers. The caging due to the mutual trapping of neighboring particles, slows down the  diffusion, and the corresponding relaxation time depends strongly on the packing of the system. The presence of periodic permanent barriers with a height that increases with increasing density,  are intrinsically associated to the layered structure of the smectic phase and determine the hopping-type layer-to-layer diffusion.   As detected in out-of-equilibrium colloidal systems, such as supercooled liquids, we found that in a given time interval some particles displace longer distances than the average, giving rise to a remarkable heterogeneous dynamics which results in significant deviations from  Gaussian behavior. We quantified these heterogeneities by computing (1) the non-Gaussian parameter $\alpha_{2, z}(t)$, which significantly deviates from zero at high volume fractions; (2) the self-part of the van Hove functions, whose long tails give clear evidence of the presence of \textit{fast} particles even at short times; (3) the mean square displacement, which exhibits a plateau quantifying the average life-time of the transient cages; and (4) the self-part of the intermediate scattering function, which characterizes the relaxation decay of the system. In the complete range of stability of the smectic phase, we observed a very slow inter-layer structural relaxation which spans over four time decades from the N-Sm to the Sm-K transition.
We note that caging as well as the permanent barrier both contribute to the non-Gaussian dynamics \cite{vorselaars}. 
The in-layer relaxation dynamics is very fast, but does not show an exponential decay at long-times, as it would be expected for simple liquids. The observed stretched-exponential decay corresponds to that of a low-density supercooled liquid where the cage effect is not sufficiently strong to yield a cage-trapping plateau.

We observed qualitative agreement with  the dynamics of freely-rotating hard rods \cite{patti}, but we do find some quantitative deviations in the non-Gaussian behavior, mainly caused by the huge differences in the height of the potential energy barriers. To be specific, the height of the barrier for aligned rods tends to zero at the continuous N-Sm transition, while it remains finite in the case of the first order N-Sm transition of freely rotating rods, which changes the dynamics significantly.

\section{Acknowledgements}
\label{sec:Acknowledgements}

This work was financed by a NWO-VICI grant.


\begin{thebibliography} {99}

\bibitem{kob}
W. Kob, C. Donati, S. J. Plimpton, P. H. Poole, and S. C. Glotzer, Phys. Rev. Lett. {\bf79}, 2827 (1997).

\bibitem{berthier}
L. Berthier, G. Biroli, J.-P. Bouchaud, L. Cipelletti, D. El Masri, D. L'H\^{o}te, F. Ladieu, and M. Pierno, Science {\bf 310}, 1797 (2005).

\bibitem{brambilla}
G. Brambilla, D. El Masri, M. Pierno, L. Berthier, and L. Cipelletti, Phys. Rev. Lett. {\bf 102}, 085703 (2009).

\bibitem{abete}
T. Abete, A. de Candia, E. Del Gado, A. Fierro, and A. Coniglio, Phys. Rev. E {\bf78}, 041404 (2008).

\bibitem{puertas}
A. M. Puertas, M. Fuchs, and M. E. Cates, J. Chem. Phys. {\bf121}, 2813 (2004).

\bibitem{richert}
R. Richert, J. Non-Cryst. Solids 172-174, 209 (1994).

\bibitem{granular}
A. Lef\`{e}vre, L. Berthier, and R. Stinchcombe, Phys. Rev. E \textbf{72} 010301(R) (2005); A. S. Keys, A. R. Abate, S. C. Glotzer, and D. J. Durian, Nat. Phys. \textbf{3}, 260 (2007).

\bibitem{lefort}
R. Lefort, D. Morineau, R. Guégan, M. Guendouz, J. Zanotti, and B. Frick, Phys. Rev. E \textbf{78} 040701(R) (2008).

\bibitem{ji}
Q. Ji, R. Lefort, R. Busselez, and D. Morineau, J. Chem Phys. \textbf{130}, 234501 (2009).

\bibitem{cang}
H. Cang, J. Li, V. N. Novikov, and M. D. Fayer, J. Chem Phys. \textbf{118}, 9303 (2003).

\bibitem{gottke}
S. D. Gottke, D. D. Brace, H. Cang, B. Bagchi, and M. D. Fayer, J. Chem Phys. \textbf{116}, 360 (2001).

\bibitem{patti}
A. Patti, D. El Masri, R. van Roij, and M. Dijkstra, arXiv:0906.3093v1

\bibitem{degennes}
P. G. de Gennes and J. Prost, The Physics of Liquid Crystals (Clarendon, Oxford, 1993).

\bibitem{frunza}
L. Frunza, S. Frunza, H. Kosslick, and A. Sch\"{o}nhals, Phys. Rev. E \textbf{78}, 051701 (2008)

\bibitem{guegan}
R. Gu\'{e}gan, D. Morineau, R. Lefort, A. Mor\'{e}ac, and W. B\'{e}ziel, J. Chem. Phys. \textbf{126}, 064902 (2007).

\bibitem{furo}
I. Fur\'{o} and S. V. Dvinskikh, Magn. Reson. Chem. {\bf 40}, S3 (2002).

\bibitem{lettinga}
M. P. Lettinga and E. Grelet, Phys. Rev. Lett. {\bf 99}, 197802 (2007).

\bibitem{bier}
M. Bier, R. van Roij, M. Dijkstra, and P. van der Schoot, Phys. Rev. Lett. {\bf 101}, 215901 (2008).

\bibitem{stroobants}
A. Stroobants, H. N. W. Lekkerkerker, and D. Frenkel, Phys. Rev. Lett. \textbf{57}, 1452 (1986).

\bibitem{veerman}
J. A. C. Veerman and D. Frenkel, Phys. Rev. A {\bf 43}, 4334 (1991).

\bibitem{Doi}
M. Doi and S. F. Edwards, The Theory of Polymer Dynamics, Clarendon Press, Oxford, 1994.

\bibitem{pryamitsyn}
V. Pryamitsyn and V. Ganesan, J. Chem. Phys. \textbf{128}, 134901 (2008).

\bibitem{vanduijneveldt}
J.S. van Duijneveldt and M.P. Allen, Mol. Phys. {\bf 90}, 243 (1997).


\bibitem{selinger}
R. L. B. Selinger, Phys. Rev. E, \textbf{65}, 051702 (2002).


\bibitem{lill}
J. V. Lill and J. Q. Broughton, Phys. Rev. B, \textbf{63}, 144102 (2001).

\bibitem{rahman}
A. Rahman, Phys. Rev. {\bf 136}, A405 (1964).

\bibitem{vorselaars}
B. Vorselaars, A.V. Lyulin, K. Karatasos, and M.A.J. Michels, Phys. Rev. E {\bf 75}, 011504 (2007).

\bibitem{hurley} M.M. Hurley and P. Harrowell, J. Chem.
Phys. {\bf 105}, 10521 (1996).
\bibitem{likos}
A.J. Moreno and C.N. Likos, Phys. Rev. Lett. {\bf 99}, 107801
(2007).



\bibitem{donati}
C. Donati, J. F. Douglas, W. Kob, S. J. Plimpton, P. H. Poole, and S. C. Glotzer, Phys. Rev. Lett. \textbf{80}, 2338 (1998).



\bibitem{weeks}
E. R. Weeks and D. A. Weitz, Phys. Rev. Lett. \textbf{89}, 095704 (2002).

\bibitem{weeks2}
Eric R. Weeks, J. C. Crocker, Andrew C. Levitt, Andrew Schofield, and D. A. Weitz, Science, \textbf{287}, 627 (2000).

\bibitem{kegel}
W. K. Kegel and A. van Blaaderen, Science \textbf{287}, 290 (2000).



\bibitem{puertas2}
A. M. Puertas, M. Fuchs, and M. E. Cates, Phys. Rev. E \textbf{67}, 031406 (2003).


\bibitem{abete2}
T. Abete, A. de Candia, E. Del Gado, A. Fierro, and A. Coniglio, Phys. Rev. Lett. \textbf{98}, 088301 (2007).

\bibitem{hansen}
J. P. Hansen and I. R. McDonald, Theory of Simple Liquids (Academic, London, 1986).

\bibitem{elmasri}
D. El Masri, G. Brambilla, M. Pierno, G. Petekidis, A. B. Schofield, L. Berthier and L. Cipelletti,  J. Stat. Mech. P07015 (2009)


\end{thebibliography}
\end{document}